# Harvesting Solar Thermal Energy With a Micro-gap Thermionic-Thermoelectric Hybrid Energy Converter: Model Development, Energy Exchange Analysis, and Performance Optimization

Ehsanur Rahman[1,2]* and Alireza Nojeh[1,2]

[1] Department of Electrical and Computer Engineering, University of British Columbia, Vancouver, BC, V6T 1Z4, Canada

[2] Quantum Matter Institute, University of British Columbia, Vancouver, BC, V6T 1Z4, Canada

*Email: ehsanece@ece.ubc.ca

**Abstract**

We present a comprehensive analysis of a dual, micro-gap thermionic-thermoelectric hybrid energy converter by developing a detailed theoretical model of the system. The Space-charge and near-field effects in thermionic conversion and the temperature-dependent effects in thermoelectric conversion are considered while studying the energy flow through the cascade system. The temperatures and energy exchange channels in the different parts of the system are quantified with a self-consistent iterative algorithm considering the energy balance condition. The model is deployed to simulate the hybrid system performance when energized by a constant heat flux source which could, for example, be concentrated solar irradiance. The effect of solar radiation intensity on the combined system efficiency, and the dependence of the thermoelectric generator performance on the waste heat released from the thermionic top stage, are illustrated with several examples. Based on these analyses, a performance optimization of the hybrid system is carried out. The findings presented in this work offer useful insights into the hybrid generator operation. Moreover, the model developed can be used as a design tool for the practical implementation of such a hybrid system, or for extracting material- and device-related parameters from experimental data by solving the reverse problem.

## 1. Introduction

Generation of electrical power using new approaches [1] has become a growing field of research due to the pressing environmental concerns as well as an increasing need to replace non-renewable sources of energy [2–4]. The realm of electrical energy conversion has long been dominated by mechanical heat engines which are complex and expensive, and involve multiple conversion stages with moving parts - a major source of energy loss in the conversion process. The heat driving these power plants can be generated using conventional fuels or it may be obtained from renewable energy sources such as solar radiation or geothermal energy. In addition, heat is also released as a byproduct of different industrial processes or everyday activities. This widespread availability of thermal energy has motivated research into various energy conversion techniques other than conventional turbine generators. Two such alternative techniques are thermionic and thermoelectric conversion [5–7]. Without any moving parts and using electrons as a working fluid, thermionic and thermoelectric converters are in principle simple devices that can be made into various form factors, be deployed both as central power stations and in off-grid scenarios, have a long lifetime, and require minimal maintenance, all of which are important advantages. Due to the different physics involved in exciting and transporting the working fluid in thermionic and thermoelectric devices, in practice the two are complementary in terms of ranges of temperature they are efficient at harvesting. Thermionic conversion requires electrons to overcome an electron emitter's work function and traverse a vacuum or plasma region to reach an electron collector and produce useful electrical power. In practice, significant thermionic emission requires a temperature of around 1000 K or higher [8,9]. On the other hand, the low barrier to electron flow in a thermoelectric converter means that it can, in theory, utilize any source of heat irrespective of the temperature. It is thus tempting to operate a thermoelectric generator at a high temperature difference between its hot and cold sides in order to maximize its thermodynamic efficiency. However, maintaining a high temperature difference in a thermoelectric generator is difficult due to heat leakage through finite lattice thermal conductivity. In addition, the Seebeck coefficient and figure of merit of practical thermoelectric materials decrease rapidly after crossing a temperature threshold [10]. This means that the high-temperature operation of a thermoelectric generator would require excessive input power without any significant gain in electrical output, thereby severely degrading the efficiency. Due to these practical concerns, thermoelectric generators are suitable for low-grade waste heat recovery [11], whereas thermionic generators can use high-temperature heat sources. This also suggests that when a high-temperature heat source is available, cascading the two technologies may yield higher efficiency than the individual thermionic and thermoelectric generators. Based on this idea, several researchers have proposed models to analyze the efficiency of a hybrid thermionic-thermoelectric generator (which, for the sake of brevity, will henceforth be termed TEC-TEG)[12,13]. The effectiveness of these models depends on the details of the device physics incorporated and the considerations of external heat transfer processes. Xuan *et al.* [12] studied a TEC-TEG by specifying the temperatures of the TEC and TEG stages; however, in reality, these temperatures are not independent and should be determined based on energy coupling between the two stages and with the environment. They also did not include the Thomson effect in the TEG stage. Wang *et al.* [13] did determine the temperatures based on energy balance; however, they did not include the space-charge effect in analyzing the TEC stage. Both works neglected near-field radiative coupling, which, through a complex interplay with the space-charge effect, has a dominant role in determining the efficiency of a TEC [14–16].Therefore, despite valuable insight gained from the above works, a comprehensive and accurate model for the TEC-TEG is missing. In addition to a self-consistent treatment of energy balance among the TEC, the TEG, and the environment, such a model would require, for the TEC stage, the analysis of interelectrode radiative coupling using fluctuational electrodynamics [17] and a charge transport analysis considering the phase space of the electrons in the gap region [8,9,15,18]. In addition, for the TEG stage, the model should consider the temperature dependence of various material parameters such as the Seebeck coefficient, electrical resistivity, and thermal conductivity. Here, we present an all-encompassing model taking the above effects into account for the TEC-TEG. This model, which is rich in the physics of both heat engines, although computationally complex, offers a powerful tool for the understanding and design of this hybrid energy converter. In addition to developing the model, we have used it to investigate the performance of a TEC-TEG based solar thermal harvesting system. This system can be deployed to convert solar thermal energy from already existing concentrated solar power (CSP) harvesting systems. For instance, such systems had a total global installed capacity of 5,500 MW in 2018 [19]. To put this into perspective, the total CSP capacity was only 354 MW in 2005. This shows that there is growing demand for CSP harvesting technology. The United States has several CSP facilities already in use such as Ivanpah Solar Power Facility (392 MW), Solar Energy Generating Systems (combined capacity of 310 MW), and Genesis Solar Energy Project (280 MW). There are also similar examples of CSP plants in other parts of the worlds.

## 2. Model Description and Computational Methodology

### 2.1. Energy balance, efficiency, and the self-consistent iterative model of a concentrated solar power-driven TEC-TEG

A schematic representation of the TEC-TEG analyzed in this work is shown in fig.1. As an illustrative example of the heat source, we consider the concentrated solar thermal system. In such a system, solar radiation incident on the earth's surface is concentrated by using mirrors of various shape (e.g. including a heliostat and a parabolic trough or dish) onto a receiver. The receiver thus heats up, resulting in solar thermal energy. The receiver can employ different mechanisms for this purpose. One approach is the use of a tube containing a liquid (also known as the solar thermal fuel) which absorbs and transfers the thermal energy to the electrical conversion system. An interesting feature of this approach is that the liquid can store the thermal energy for later conversion to electricity. Another approach is to deliver the concentrated solar power directly to the converter by using a selective absorber. (A selective absorber is a material with carefully-designed spectral emissivity (or absorptivity) [20–23]). We emphasize that our model will be equally valid for other types of heat source, too. The heat flux from the concentrated solar thermal system, $Q_{\mathrm{Solar}}$, is absorbed by the emitter (also known as the cathode) electrode of the TEC, heating the electrons and the lattice. Some of the electrons gain sufficient energy to overcome the emitter vacuum barrier and are emitted into the interelectrode space. The energy flux density carried by these thermionically emitted electrons is denoted as $Q_{\mathrm{T}}$ . Due to the temperature difference between emitter and collector (also known as anode), a part of the incident energy flux is exchanged between the two electrodes radiatively, the density of which we denote by $Q_{\mathrm{Rad}}$.

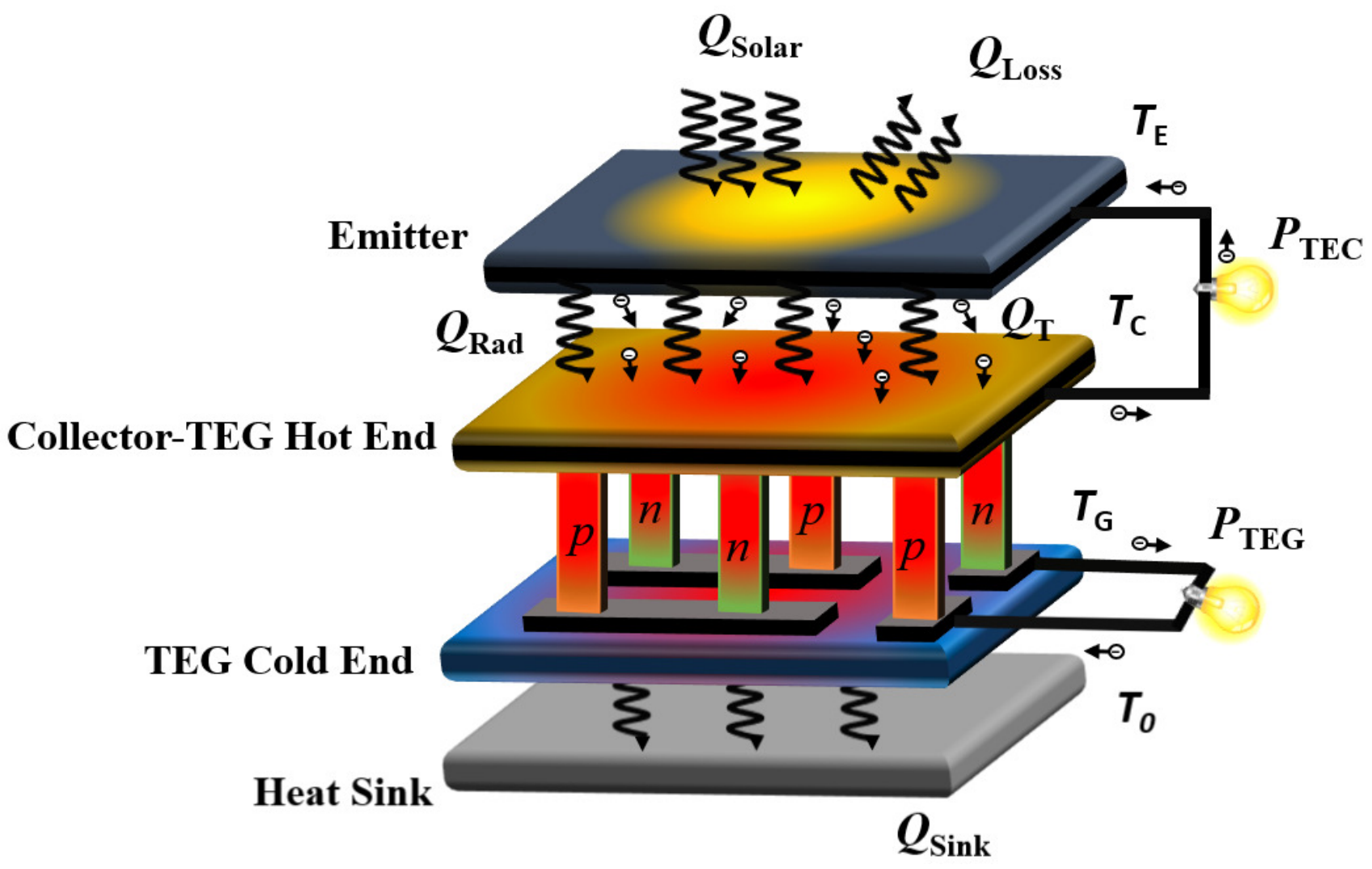

Fig. 1. The input, intermediate, and output energy fluxes in a concentrated solar-powered TEC-TEG system. The symbols *n* and *p* represent the n-type and p-type semiconductor legs of the thermocouple, respectively.

This interelectrode radiative heat transfer consists of contributions from propagating waves and evanescent waves when the interelectrode gap is a few micrometers or less (which is required to mitigate the space-charge effect and obtain high efficiency [24–26]). A part of the incident energy is also lost as radiation from the emitter to the ambient and is given by $Q_{\mathrm{Loss}} = \varepsilon\sigma(T_{\mathrm{E}}^{4} - T_{0}^{4})$, where $T_E$ and $T_0$ are the TEC emitter and ambient temperature, respectively, $\varepsilon$ is the effective emissivity of the emitter system and $\sigma$ is the Stefan Boltzmann constant. In steady-state, the energy flux input to the emitter is equal to the sum of the fluxes leaving from the emitter to the collector and the surroundings:

$$Q_{\text{Solar}} = Q_{\text{T}} + Q_{\text{Rad}} + Q_{\text{Loss}} \quad . \qquad (1)$$

A portion of the thermionic energy flux transferred from the emitter to the collector is converted into useful electrical output. If $J_{\text{TEC}}$ is the net current density from the emitter to the collector and $V_{\text{TEC}}$ is the operating voltage, then the output power density and efficiency of the TEC can be defined, respectively, as

$$P_{\text{TEC}} = J_{\text{TEC}} V_{\text{TEC}} \;\; (2.\text{a}) \quad \text{and} \quad \eta_{\text{TEC}} = \frac{P_{\text{TEC}}}{Q_{\text{Solar}}} \quad . \quad (2.\text{b})$$

The rest of the energy is delivered to the collector as heat. The corresponding energy flux density, which we call $Q_{\text{H}}$, can thus be written as

$$Q_{\text{H}} = Q_{\text{T}} + Q_{\text{Rad}} - P_{\text{TEC}} . \qquad (3)$$

This is the input to the TEG. In our hybrid system, the TEG is physically connected to the TEC upper stage. In a TEG, the modules (which are p and n-type semiconductor pairs) are connected thermally in parallel between a hot and a cold plate, which are thermally conductive but electrically insulating. In our hybrid device, the hot side of the TEG is in direct thermal contact with the collector of the TEC. A part of $Q_{\text{H}}$ is converted to useful electrical power with a power density, $P_{\text{TEG}}$, which is then fed to a separate load with an independent return circuit. The rest of the heat is released from the TEG's cold end to a heat sink and this heat transfer process can be expressed by

$$Q_{\text{Sink}} = K_{\text{L}}(T_{\text{G}} - T_{0}), \qquad (4)$$

where $K_{\text{L}}$ is the thermal conductance per unit area between the collector and the heat sink and $T_{\text{G}}$ is the temperature of the TEG cold side. The conversion efficiencies of the TEG and the TEC-TEG can be defined, respectively, as

$$\eta_{\text{TEG}} = \frac{P_{\text{TEG}}}{Q_{\text{H}}} \;\; (5.\text{a}) \quad \text{and} \quad \eta_{\text{Combined}} = \frac{P_{\text{TEC}} + P_{\text{TEG}}}{Q_{\text{Solar}}} . \qquad (5.\text{b})$$

A flow chart for the self-consistent iterative algorithm to implement the above model considering energy balance at the different electrodes is shown in fig. 2. The self-consistent algorithm developed in this work is used to calculate the temperature at different parts of the hybrid device as well as the different energy exchange channels. The algorithm takes material- and device-related parameters as input at the start of the iterative solution process. To start the iteration, an initial guess of temperatures is provided. With this initial guess, the algorithm then checks for convergence of the energy balance criterion at different parts of the hybrid device and updates the temperatures accordingly until convergence is achieved. Here we note that the various energy exchange channels such as thermionic and radiative heat flux have strong nonlinear dependences on the electrode temperatures. Therefore, adaptively updated coefficients have been used to update the electrode temperatures during the self-consistent cycles.

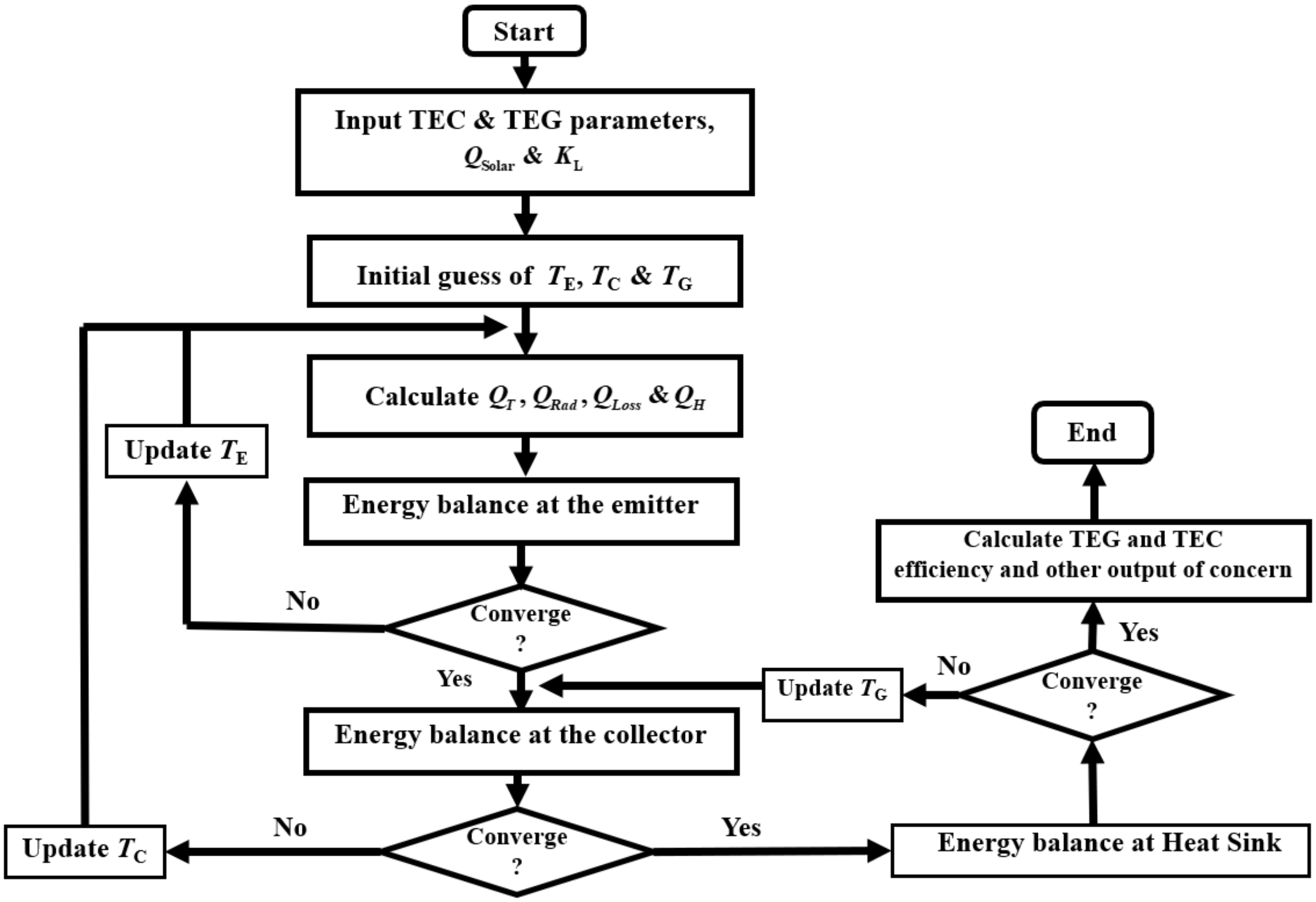

Fig. 2. Flowchart of the self-consistent algorithm used to implement the model.

## 2.2. The Space-charge effect in the interelectrode region of a TEC

The space-charge effect in a TEC arises from the coulombic repulsion caused by the electrons which are in transit in the interelectrode space. The energy diagram of a space-charge limited TEC is shown in fig. 3. As can be seen, an energy barrier is formed in the gap when the interelectrode electron concentration is very high (which is typically the case for macroscopic gaps). Due to this barrier, only a portion of the electrons overcoming the emitter work function, which have sufficient energy to also surpass this additional barrier, can reach the collector; lower energy electrons are reflected back to the emitter. This phenomenon significantly reduces the output current density. In this work, the space-charge effect is derived assuming that electrons traversing the interelectrode distance are collisionless particles [8,9]. The detailed implementation of the theory has been outlined in the supplementary document and will not be repeated here. In brief, the emitter and collector current of a TEC can be defined, respectively, as

$$J_{\mathrm{E}} = A_0 T_{\mathrm{E}}^{\,2} e^{\left(-\frac{\varphi_{\mathrm{m}}}{k_{\mathrm{B}} T_{\mathrm{E}}}\right)} \quad (6.a) \quad \text{and} \quad J_{\mathrm{C}} = A_0 T_{\mathrm{C}}^{\,2} e^{\left(-\frac{\varphi_{\mathrm{m}} - eV_{\mathrm{TEC}}}{k_{\mathrm{B}} T_{C}}\right)}, \quad (6.b)$$

where $A_0$ is the Richardson-Dushmann Constant and $\varphi_{\mathrm{m}}$ is the maximum motive in the interelectrode space.

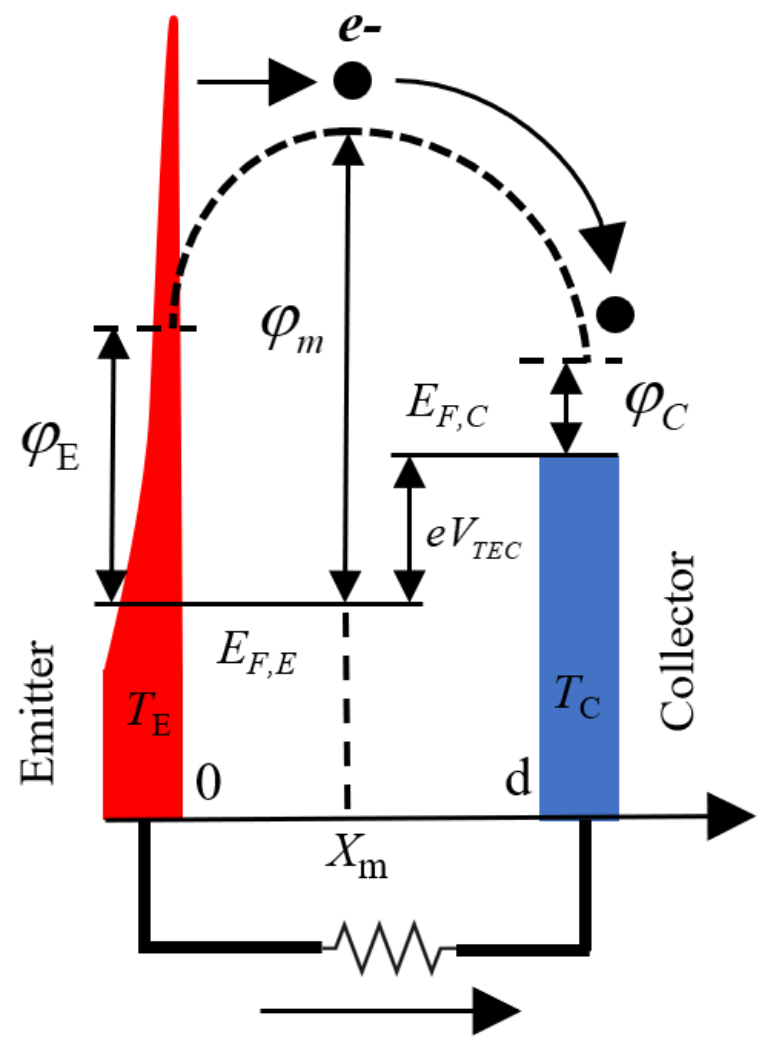

Fig. 3. The energy diagram of a TEC in the space-charge regime. $E_{F,E}$ and $E_{F,C}$ are the Fermi levels of the emitter and the collector, respectively. $\varphi_m$ is the maximum motive in the interelectrode space and $e$ is the electron charge. $\varphi_E$ and $\varphi_C$ are the work functions of the emitter and the collector, respectively. $T_E$ and $T_C$ are the temperatures of the emitter and the collector, respectively. $V_{TEC}$ is the voltage difference between the two electrodes. $x_m$ is the position of the maximum motive and d is the interelectrode gap width.

The value of the maximum motive, $\varphi_m$, will depend on the interelectrode distance and operating voltage of the TEC. Depending on $\varphi_m$, the TEC will operate in saturation, space-charge or retarding modes. The details of the phase space analysis and the expression for $\varphi_m$ in different modes of the TEC operation are provided in the supplementary document.

Knowing the current density at the two electrodes of a TEC, the net energy flux carried by the thermionic current from the emitter is given by [27]

$$Q_T = \frac{[(J_E - J_C)\varphi_m + 2k_B(T_E J_E - T_C J_C)]}{e} . \quad (7)$$

A part of this thermionic energy flux is converted to electricity while the rest is deposited in the collector as heat when thermionic electrons are absorbed by it. The waste heat flux which is fed to the TEG hot side from the TEC collector is given by

$$Q_H = \frac{[(J_E - J_C)(\varphi_m - eV_{TEC}) + 2k_B(T_E J_E - T_C J_C)]}{e} + Q_{Rad} . \quad (8)$$

### 2.3. Radiative heat transfer between emitter and collector of a TEC

When the interelectrode distance in a thermionic device is large, radiative heat exchange between the electrodes is due to the far-field propagating waves and is given by the Stefan-Boltzmann law. However, mitigating the space-charge effect requires the electrodes to be placed at a distance of a few micrometers or less. This is on the order of the characteristic wavelength of thermal radiation from the emitter, which is given by Wien's displacement law as $\lambda_T = 2.9x10^{-3}$ m.K / $T_E$, and so the emitter and collector surfaces are coupled by evanescent waves. The coupling of these evanescent waves significantly enhances the radiative heat transfer in the near-field regime. The energy transfer including both propagating and evanescent components can be modelled using fluctuational electrodynamics [28–30] as

$$Q_{\text{prop}} = \frac{1}{\pi^2}\int_0^\infty dw\,[\Theta_{\text{E}}(w,T_{\text{E}}) - \Theta_{\text{C}}(w,T_{\text{C}})] \times \int_0^{\frac{w}{c}} S_{\text{prop}}(w,\beta,\varepsilon_{\text{E}},\varepsilon_{\text{C}})\,d\beta \tag{9}$$

and

$$Q_{\text{evan}} = \frac{1}{\pi^2}\int_0^\infty dw\,[\Theta_{\text{E}}(w,T_{\text{E}}) - \Theta_{\text{C}}(w,T_{\text{C}})] \times \int_{\frac{w}{c}}^{\infty} S_{\text{evan}}(w,\beta,\varepsilon_{\text{E}},\varepsilon_{\text{C}})\,d\beta \tag{10}$$

, where $Q_{\text{prop}}$ and $Q_{\text{evan}}$ are the radiative fluxes due to propagating and evanescent waves, respectively and $\Theta(w,T)$ is the mean energy of a Planck oscillator at an angular frequency $w$ and temperature $T$ [30]. In the above equations, $\beta$ is the wavevector component parallel to the interface, $c$ is the speed of light, $\hbar$ is the reduced Planck constant, and $S_{\text{prop}}$ , $S_{\text{evan}}$ are the coupling coefficients for the propagating and evanescent waves, respectively. The contributions of the propagating and evanescent waves in heat exchange between two semi-infinite tungsten plates (which is considered as the material of TEC electrodes in this work) are explained and shown in detail in the supplementary document.

### 2.4. The power output and efficiency of a TEG

The operation of the TEG is modelled considering Peltier heating at the hot and cold junctions, Fourier thermal conduction through the legs, Joule heating because of the electrical resistance of the legs, and Thomson heating due to the temperature gradient in the legs. (While some works have neglected the Thomson effect [31,32], it has been reported that that would overestimate the performance of the TEG [33,34].) According to the energy balance condition, the heat flux absorbed by the TEG hot end from the thermionic collector should equal that leaving it. Therefore, we have [35]

$$A_{\text{TEG}}\ Q_{\text{H}} = N[\alpha_{\text{h}} T_{\text{C}} I + K(T_{\text{C}} - T_{\text{G}}) - \frac{1}{2}\overline{\mu} I (T_{\text{C}} - T_{\text{G}}) - \frac{1}{2} I^2 R] \qquad (11)$$

, where $\alpha_{\text{h}}, \overline{\mu}, K$ and $R$ are the total Seebeck coefficient at the hot side, total Thomson coefficient, total thermal conductance and total internal resistance of a p-n thermoelectric pair, respectively. $N$ is the total number of such pairs in the TEG, $A_{\text{TEG}}$ is the total cross-sectional area and $I$ is the total electrical current through the TEG. The expressions for $\alpha_{\text{h}}$, $\overline{\mu}, K$ and $R$ can be given as

$$\alpha_{\text{h}} = (\alpha_{\text{hp}} - \alpha_{\text{hn}}) \ , \qquad (12)$$

$$\overline{\mu} = (\overline{\mu}_{\text{p}} - \overline{\mu}_{\text{n}}) \quad , \qquad (13)$$

$$K = (\frac{\overline{k}_{\text{p}} A_{\text{p}}}{l_{\text{p}}} + \frac{\overline{k}_{\text{n}} A_{\text{n}}}{l_n}) \ , \qquad (14)$$

and $$R = (\frac{\overline{\rho}_{\text{p}} l_{\text{p}}}{A_{\text{p}}} + \frac{\overline{p}_{\text{n}} l_n}{A_{\text{n}}}) \ . \qquad (15)$$

In the above equations, $\alpha_{\text{hp}}, \overline{\mu}_{\text{p}}, \overline{k}_{\text{p}}, \overline{\rho}_{\text{p}}, A_{\text{p}}$ and $l_{\text{p}}$ are the hot side Seebeck coefficient, Thomson coefficient, thermal conductivity, electrical resistivity, area and length of the p-type leg, respectively. The symbols with subscript n represent the corresponding parameters of the n-type leg. We assume that the n-type and p-type legs are both structurally and characteristically symmetric except having an opposite sign of the Seebeck coefficient. The symbols with a bar represent the value of the corresponding parameter at an average temperature, $T_{\text{avg}} = (T_{\text{C}} + T_{\text{G}})/2$, where $T_{\text{C}}$ and $T_{\text{G}}$ are the temperatures of the TEG hot side and cold side, respectively.

The total heat released to the heat sink from the TEG cold side (which, according to energy balance, is the same as $Q_{\text{Sink}}$ ) equals that absorbed by it, which is given by

$$A_{\text{TEG}}\ Q_{\text{L}} = N[\alpha_{\text{c}} T_{\text{G}} I + K(T_{\text{C}} - T_{\text{G}}) + \frac{1}{2}\overline{\mu} I (T_{\text{C}} - T_{\text{G}}) + \frac{1}{2} I^2 R] . \qquad (16)$$

In the above equation, $\alpha_{\mathrm{c}} = (\alpha_{\mathrm{cp}} - \alpha_{\mathrm{cn}})$ is the total Seebeck coefficient at the cold side of a p-n pair; $\alpha_{\mathrm{cp}}$ and $\alpha_{\mathrm{cn}}$ are the cold side Seebeck coefficients of the individual p-type and n-type legs, respectively. See the supplementary document for the derivation of the above energy balance equations at the TEG hot and cold end.

The output power of the TEG can be derived from the energy balance condition as

$$A_{\mathrm{TEG}}\ P_{\mathrm{TEG}} = A_{\mathrm{TEG}}\ (Q_{\mathrm{H}} - Q_{\mathrm{L}}) = N[(\alpha_{\mathrm{h}} T_{\mathrm{C}} - \alpha_{\mathrm{c}} T_{\mathrm{G}})I - \bar{\mu} I (T_{\mathrm{C}} - T_{\mathrm{G}}) - I^2 R] = I^2 R_{\mathrm{L}} \quad (17)$$

, where $R_{\mathrm{L}}$ is the external load resistance connected to the TEG output terminals.

From the above relation, the TEG electrical current $I$ can be expressed as

$$I = \frac{[(\alpha_{\mathrm{h}} T_{\mathrm{C}} - \alpha_{\mathrm{c}} T_{\mathrm{G}}) - \bar{\mu}(T_{\mathrm{C}} - T_{\mathrm{G}})]}{R + \frac{R_{\mathrm{L}}}{N}} \ . \quad (18)$$

It can be shown by setting the derivative of $I^2 R_{\mathrm{L}}$ ,with respect to $R_{\mathrm{L}}$ , to zero that $R_{\mathrm{L}} = NR$ will maximize the power delivered to the load.

The conversion efficiency of the TEG can be given by

$$\eta_{TEG} = \frac{P_{\mathrm{TEG}}}{Q_{\mathrm{H}}} = \frac{I^2 R_{\mathrm{L}}}{N[\alpha_{\mathrm{h}} T_{\mathrm{C}} I + K(T_{\mathrm{C}} - T_{\mathrm{G}}) - \frac{1}{2}\bar{\mu} I (T_{\mathrm{C}} - T_{\mathrm{G}}) - \frac{1}{2} I^2 R]} \ . \quad (19)$$

To accurately predict the performance of a TEG, it is crucial to know the temperature dependence of the Seebeck coefficient, thermal conductivity, and electrical resistivity of the materials. For our analysis, we choose the material properties of $Bi_2Te_3$ which is widely used in the commercial thermoelectric applications and has well-defined relations of the above-mentioned material properties with temperature as follows [12,35]:

$$\alpha = (22224 + 930.6T - 0.9905T^2)\ 10^{-9}\ \mathrm{VK^{-1}} \ , \quad (20)$$

$$\rho = (5112 + 163.4T + 0.6279T^2)\ 10^{-10}\ \Omega\mathrm{m}\ , \qquad (21)$$

$$k = (62605 - 277.7T + 0.4131T^2)\ 10^{-4}\ \mathrm{Wm^{-1}K^{-1}}\ , \qquad (22)$$

where $\alpha$, $\rho$ and $k$ are the Seebeck coefficient, electrical resistivity, and thermal conductivity of the material, respectively. The Thomson coefficient is given by the Kelvin relationship,

$$\mu = T\frac{d\alpha}{dT} \quad . \qquad (23)$$

## 3. Results and Discussion

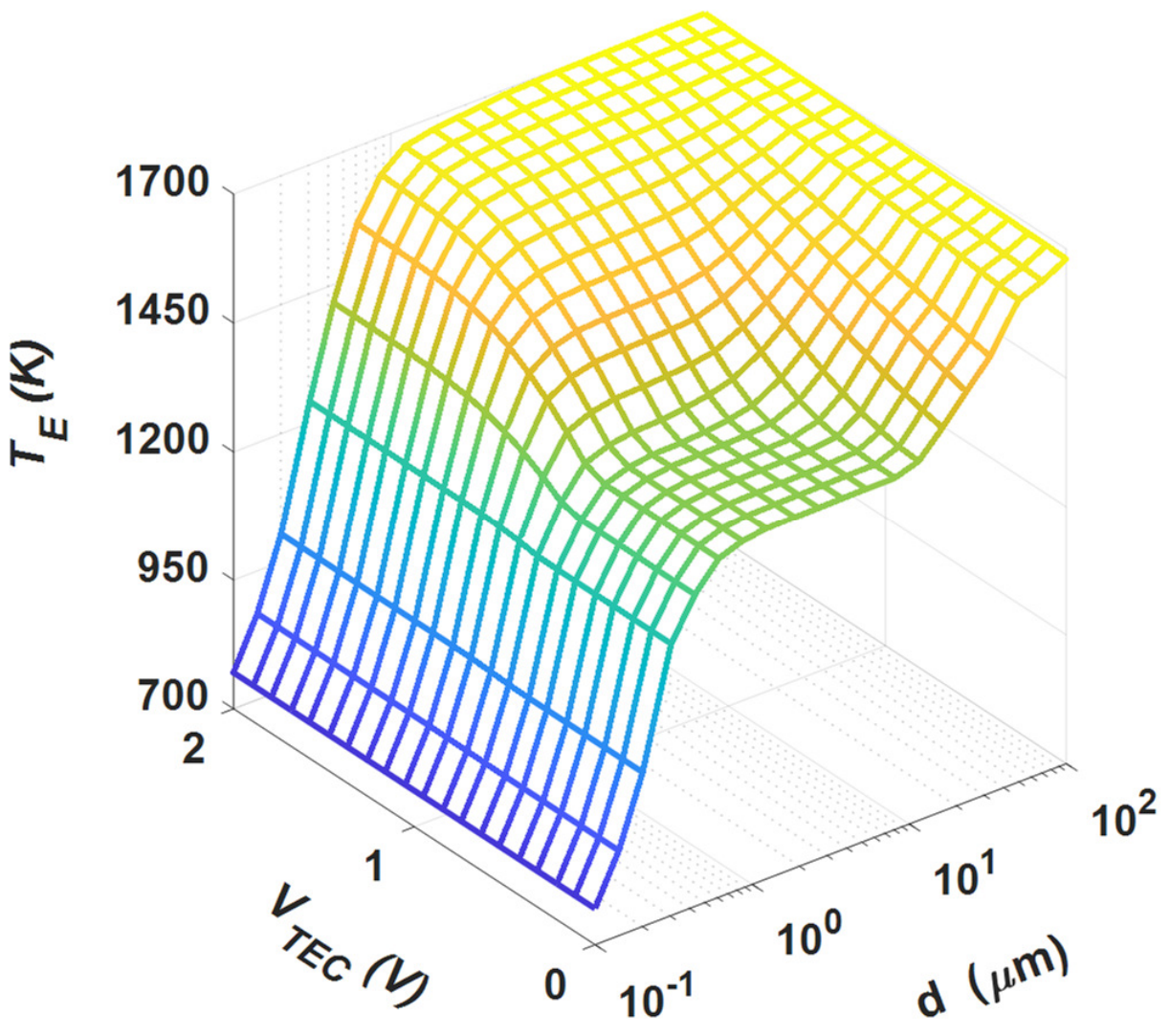

**(a)**

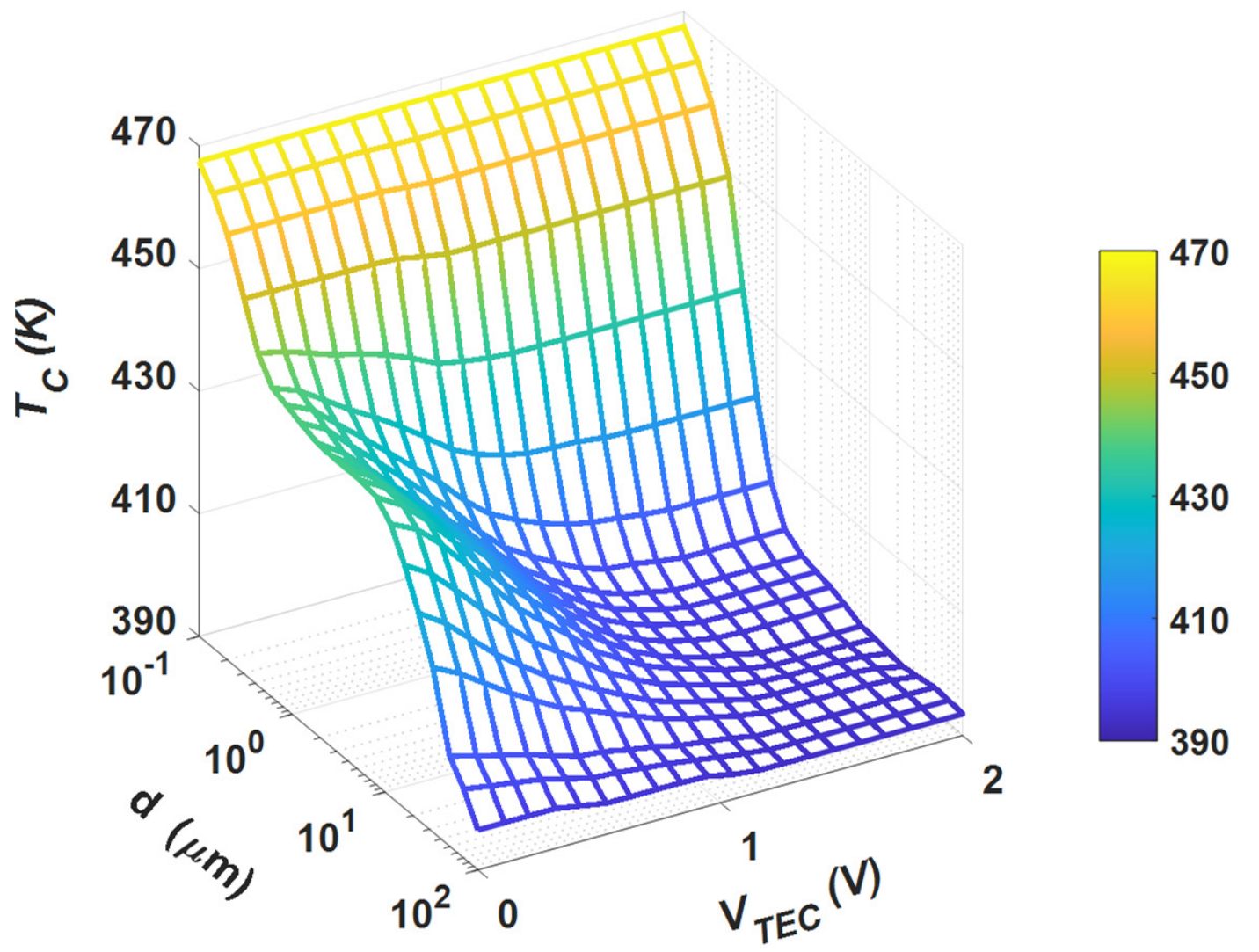

**(b)**

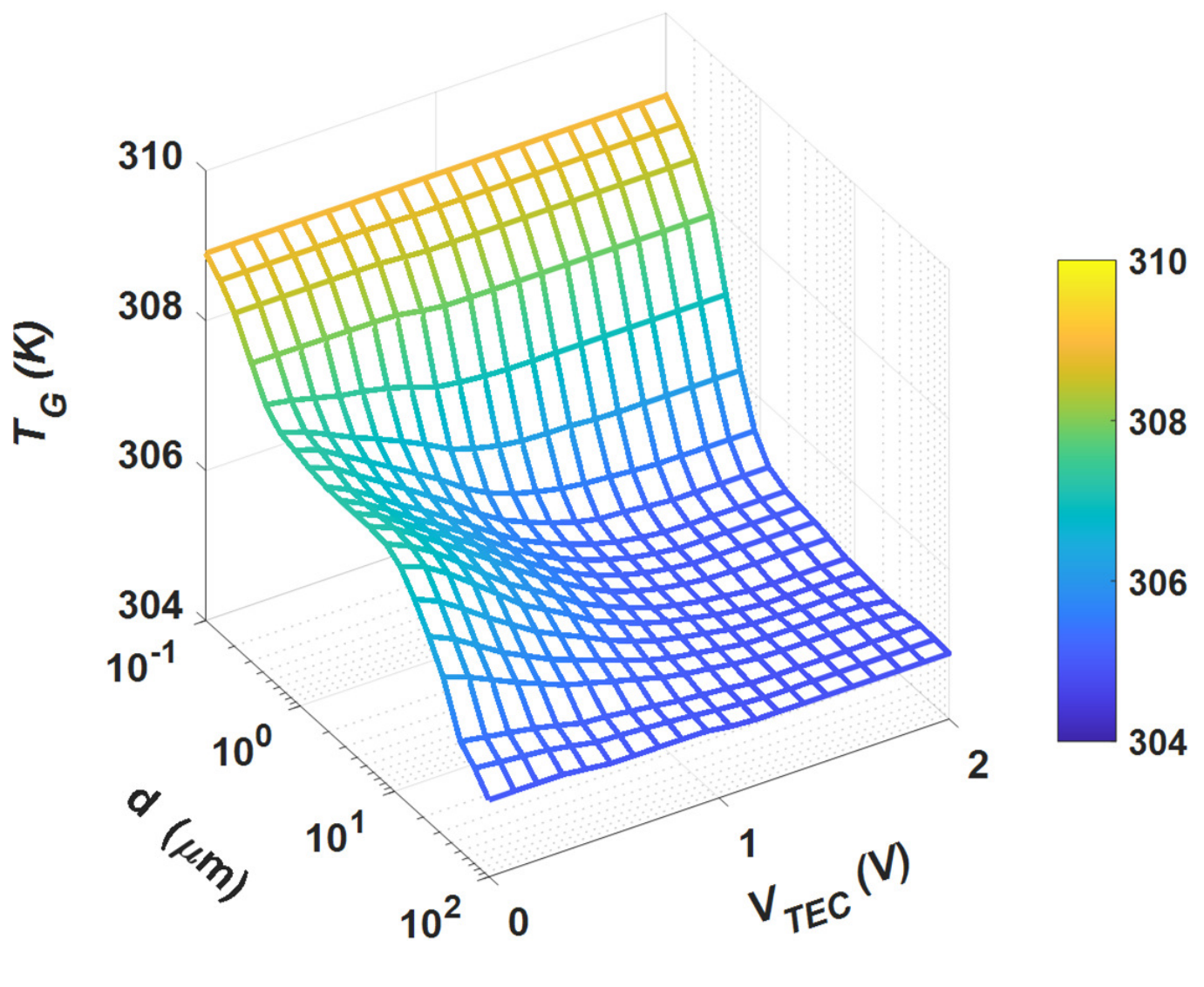

**(c)**

Fig. 4. Temperature graphs for (a) TEC emitter, (b) TEC collector, and (c) TEG cold end as a function of TEC output voltage and gap width. The graphs are shown for AM 1.5 solar radiation when concentrated by a factor of the order of 100. Note the different orientation of part (a) compared to parts (b) and (c) for better visibility.

The various energy exchange channels and electrical outputs in thermionic and thermoelectric generators depend to different extents on the temperatures of their respective electrodes. For example, in a TEC, the heat carried away from the emitter by thermionically emitted electrons has a strong nonlinear dependence, including an exponential behaviour, on the emitter temperature. On the other hand, the interelectrode radiation exchange has a fourth-power dependence on the electrode temperatures in the far-field region. As well, given that the material parameters of a TEG depend on temperature, the energy fluxes in the TEG also have nonlinear dependencies on temperatures of the hot and cold sides. Considering these factors, it is instructive to study the temperatures at different electrodes of the TEC-TEG hybrid system. In the TEC-TEG, heat is received from the external source by the TEC, which partly converts it into electrical output; it releases the rest as heat which drives the TEG stage. The amount of this heat and, therefore, the TEG operation crucially depends on the TEC gap width and operating voltage.

To study this strong dependence, fig. 4 shows the electrode temperatures as a function of the TEC gap width and output voltage over ranges of 0.1 μm to 100 μm and 0 to 2 V, respectively. The AM 1.5 solar radiation concentrated by a factor of the order of 100 has been used as a source of heat. The emitter and collector work functions of the TEC are chosen as $\varphi_E$ =2.2 eV and $\varphi_C$ =1.5 eV, respectively. The choice of these work functions is not arbitrary, but is based on the combined requirements of both high emission current and high output voltage. For the Richardson constant, we have used the universal value of 120 A $cm^{-2}k^{-2}$. An effective emissivity of 0.1 has been assumed for the radiation loss from TEC emitter to the ambient and a thermal conductance of 1 $Wcm^{-2}$ has been considered for the heat flow between the TEG cold end and the heat sink. It can be seen from fig. 4(a) and (b) that the TEC emitter and collector temperatures have negligible dependence on the TEC output voltage when the gap width is very small or large. This is because, when the gap is too small, near-field radiative exchange dominates; when the gap is too large, space-charge limits the thermionic current. Therefore, in both these limits, it is primarily radiative exchange that determines the temperatures. In between these two regions, the emitter and collector temperatures have diverse trends with the TEC voltage and gap width, arising from the interplay between the thermionic and radiative energy exchange channels as we have explained in detail elsewhere [15]. Due to the diffusive nature of heat flow in the TEG legs, the temperature on the cold end of the TEG (fig. 4(c)) follows a trend similar to $T_C$, although with much smaller variations given the good thermal contact with the heat sink.

The trends in the output power (fig. 5) of the two stages can be understood based on the above trends in electrode temperatures. In fig. 5(a), it can be seen that the TEC output power initially increases, then reaches a maximum and finally decreases as a function of both TEC voltage and gap width. The low TEC output power, at small gaps, is due to the low emitter temperature (see fig. 4(a)) caused by a strong near-field radiative exchange with the collector. On the other hand, at large gaps, the space-charge effect will severely limit the thermionic current and hence the output power. With an increase in TEC voltage, the output power initially grows; however, beyond a certain point, rapid reduction in current with a further increase of the voltage will lead to the decrease of the output power.

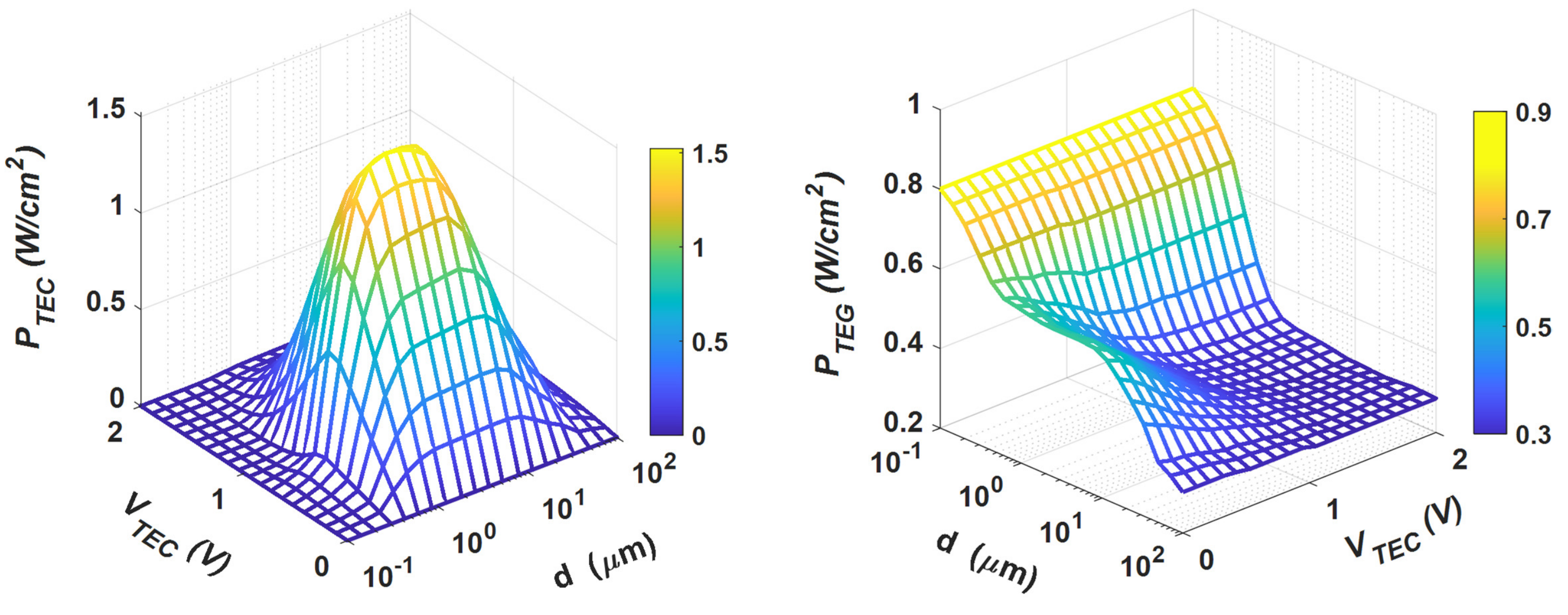

(a) (b)

Fig. 5. Output power graphs for (a) TEC and (b) TEG in the combined device as a function of TEC output voltage and gap width. The graphs are shown for AM 1.5 solar radiation when concentrated by a factor of the order of 100. Note the different orientation of part (a) compared to part (b) for better visibility.

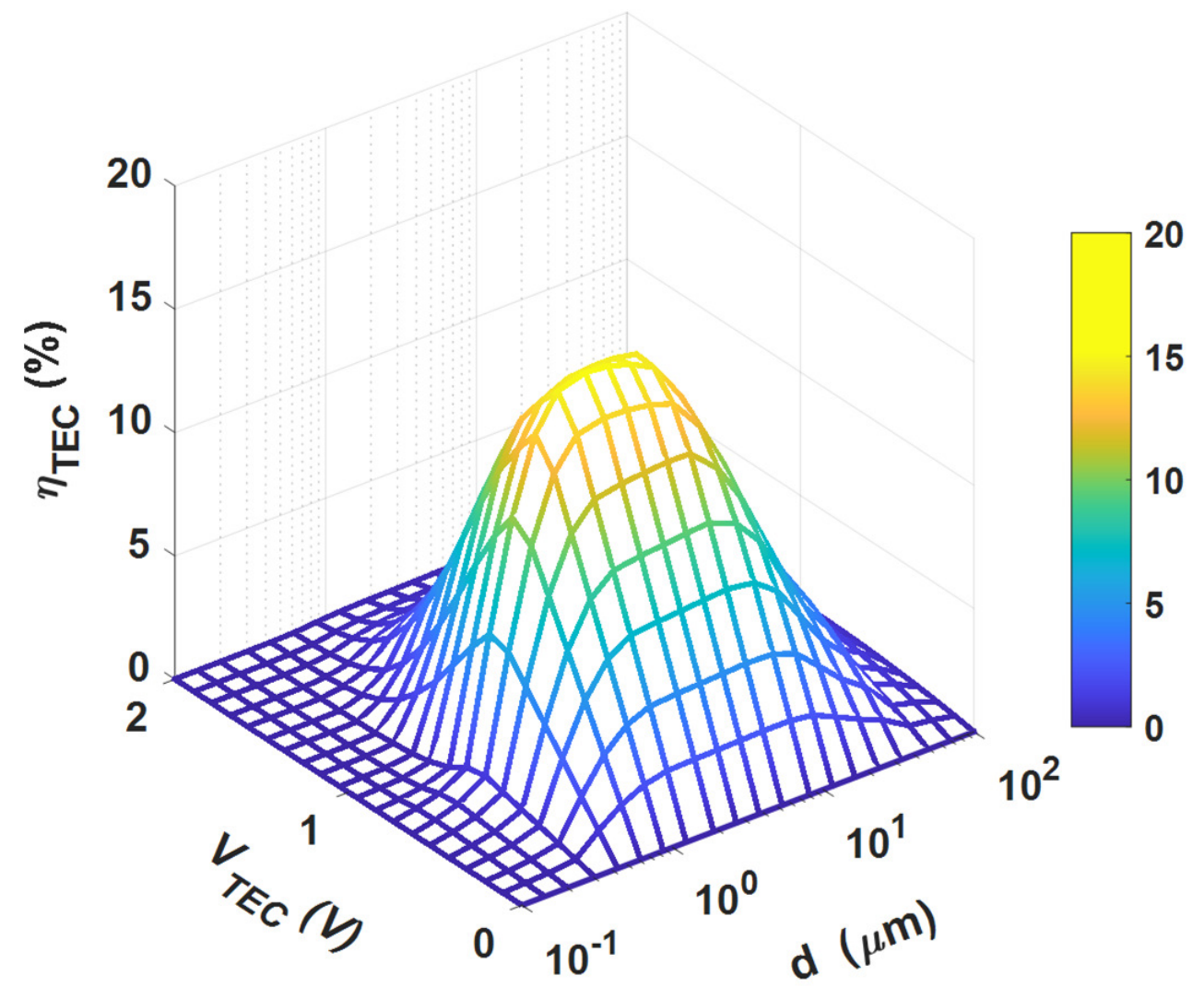

**(a)**

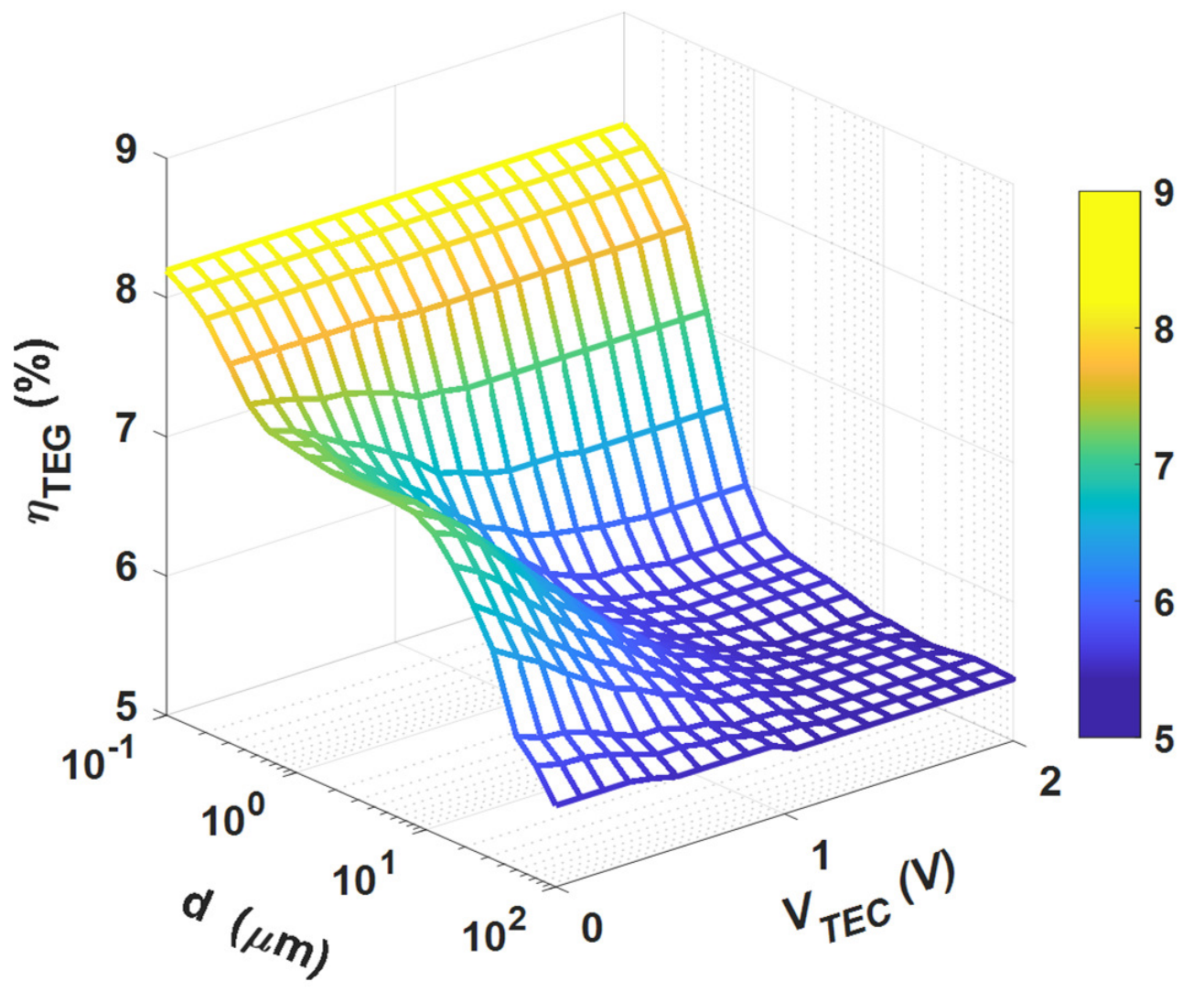

**(b)**

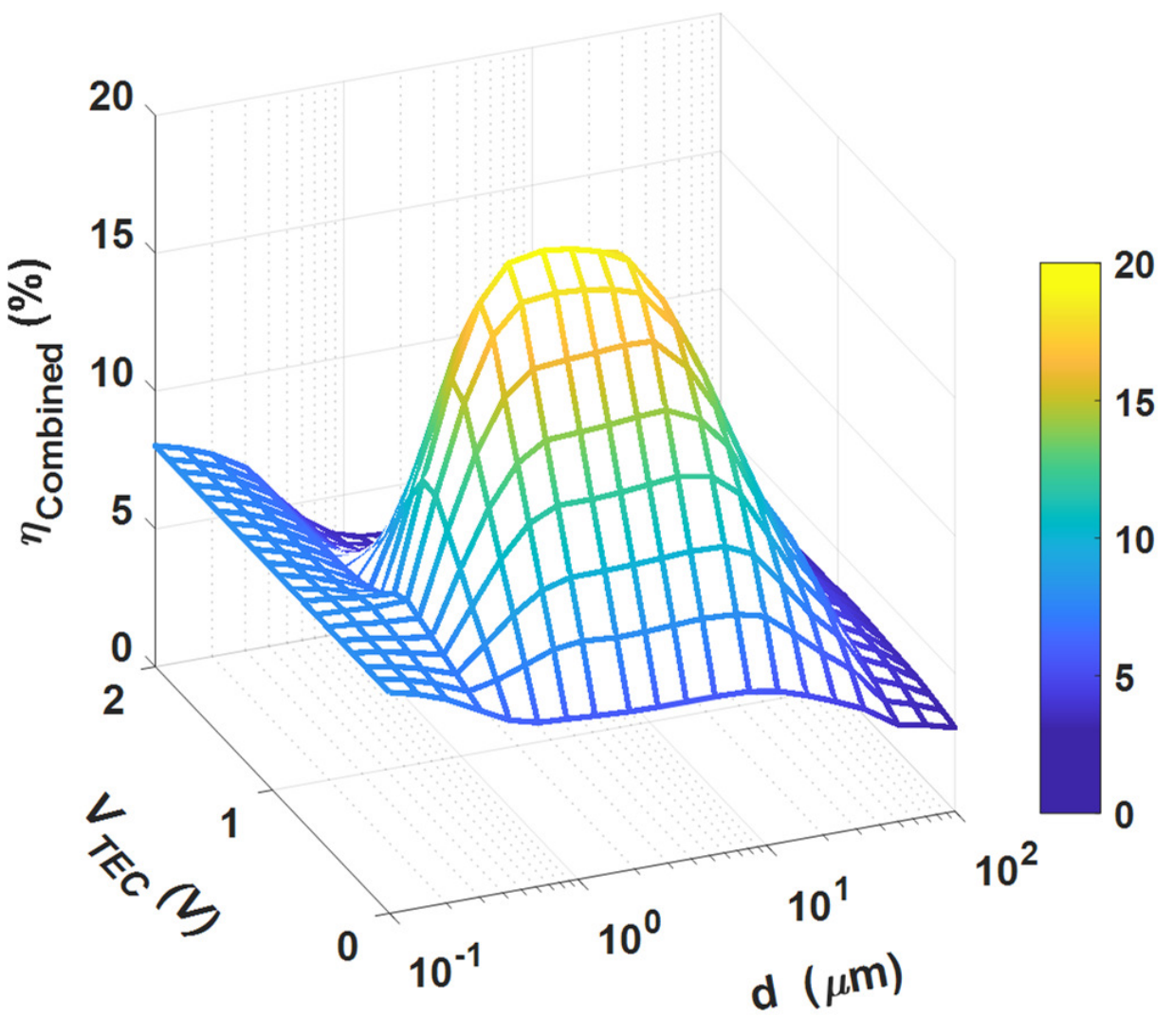

**(c)**

Fig. 6. Efficiency graphs for (a) TEC and (b) TEG in the combined device-as a function of TEC output voltage and gap width, as well as (c) total efficiency of the hybrid system. The graphs are shown for AM 1.5 solar radiation when concentrated by a factor of the order of 100. Note the different orientation of part (b) compared to parts (a) and (c) for better visibility.

As will be discussed later, for a TEG, at relatively low operating temperatures, power density and efficiency increase with an increase in the average temperature of the hot and cold sides. Therefore, given the relatively stable temperature of the cold side (because of good thermal contact with the heat sink), the output power of the TEG shown in fig. 5(b) follows a trend similar to the temperature of its hot side (which happens to be that of the TEC collector due to the thermal contact between TEC and TEG).

The conversion efficiencies of the TEC and TEG stages and the combined system are shown in fig. 6 as a function of TEC voltage and gap width. We note that the combined-system efficiency is dominated by that of the TEC at intermediate gap widths and by that of the TEG at small and large gaps. This point is further illustrated in fig.7(a) where, at each gap width, the peak efficiency (that is, the highest value of efficiency as a function of $V_{\mathrm{TEC}}$) of the combined device is shown, as well as the efficiencies of the TEC and TEG stages at that same value of $V_{\mathrm{TEC}}$.

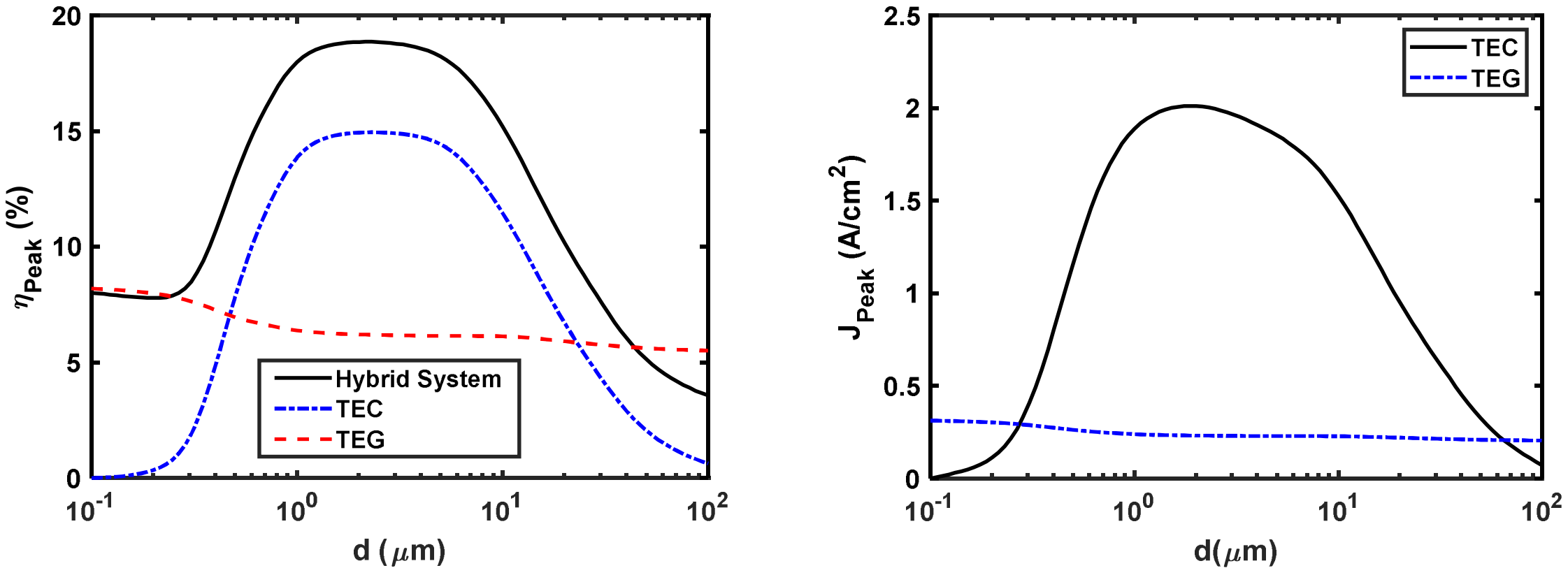

**(a)** **(b)**

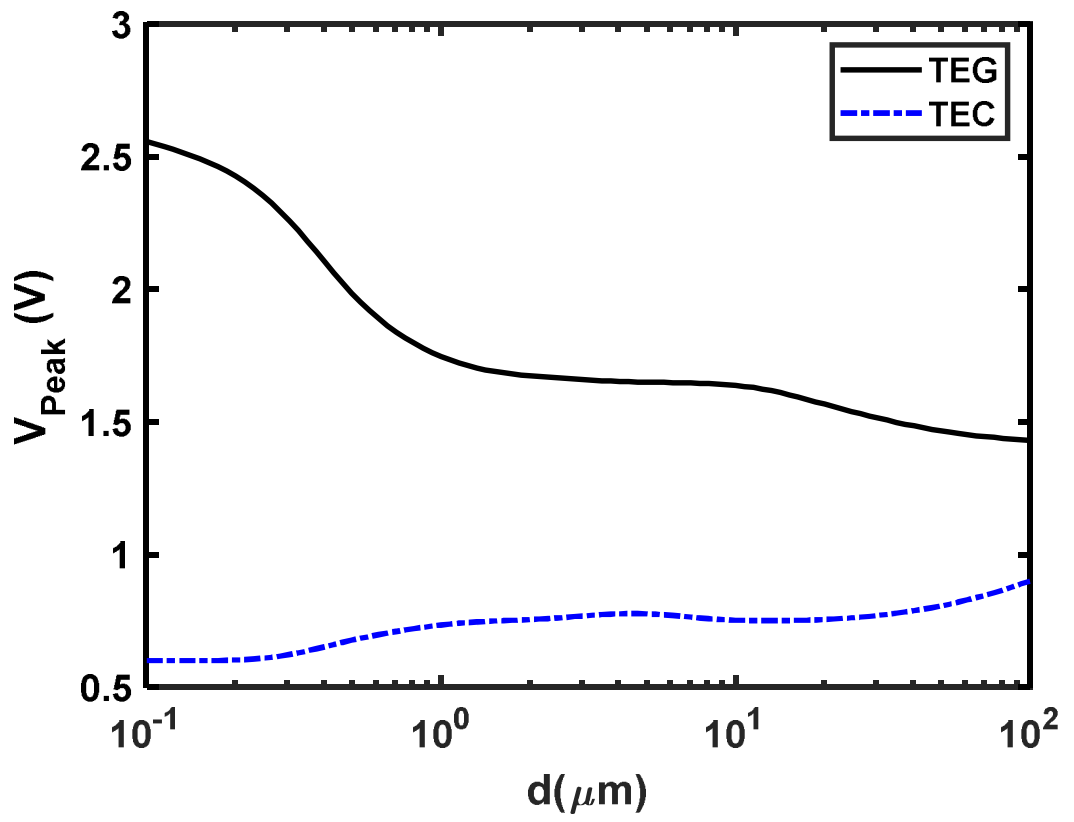

**(c)**

Fig. 7. The peak efficiency (that is, the highest value of efficiency as a function of $V_{TEC}$) of the combined device, as well as the efficiencies of the TEC and TEG stages at that same value of $V_{TEC}$, as a function of TEC gap width. (b) The output current density of the TEC and TEG stage at the peak efficiency point as a function of the TEC gap width. (c) The output voltage of the TEC and TEG stage at the peak efficiency point as a function of the TEC gap width. The graphs are shown for AM 1.5 solar radiation when concentrated by a factor of the order of 100.

The dependence of the TEC and TEG output current density on the TEC gap width at the peak efficiency of the hybrid system is shown in fig. 7(b). The current density trend of the TEC stage shows a local maximum which is again due to the near-field effect at small distances and space-charge effect at large distances. The TEG current density is relatively insensitive to the gap width variation. Fig. 7(c) shows the output voltage variation of the TEC and TEG stage at the peak efficiency with TEC gap width. It can be seen that the TEC output voltage increases with the gap width (as the TEC is gradually driven into the space charge region with an increasing interelectrode distance) while the TEG output voltage decreases. However, the TEG output voltage might also show a local maximum at higher solar intensity. This is due to the temperature dependent Seebeck coefficient of practical thermoelectric materials, which also shows a peak value at a particular temperature. These variations of TEG efficiency, current density and voltage output again indicate the significance of considering the temperature dependence of thermoelectric material parameters in modelling the TEG operation.

At this point, it is worth mentioning that the peak efficiency trend vs gap width for the TEG stage, shown in fig. 7(a), is not unique but depends on the intensity of the solar radiation. Depending on the solar concentration factor, three distinct trends can be observed, as shown in fig. 8. As can be seen from fig. 8, when the solar intensity is relatively low (*e.g.* a concentration factor of the order of 100), the TEG efficiency decreases monotonically with the gap width. For a medium solar intensity (*e.g.* a concentration factor of the order of 400), the TEG efficiency shows an upward trend with the gap width. On the other hand, for a higher solar intensity (*e.g.* a concentration factor of the order of 700), the trend exhibits a local maximum. These trends are due to the strong temperature dependence of TEG efficiency (which is again due to the temperature dependence of the material properties, as discussed before), which peaks at a particular average temperature of the hot and cold ends, and the fact that TEG hot and cold end temperatures also depend on the TEC gap width and incident solar radiation intensity [15].

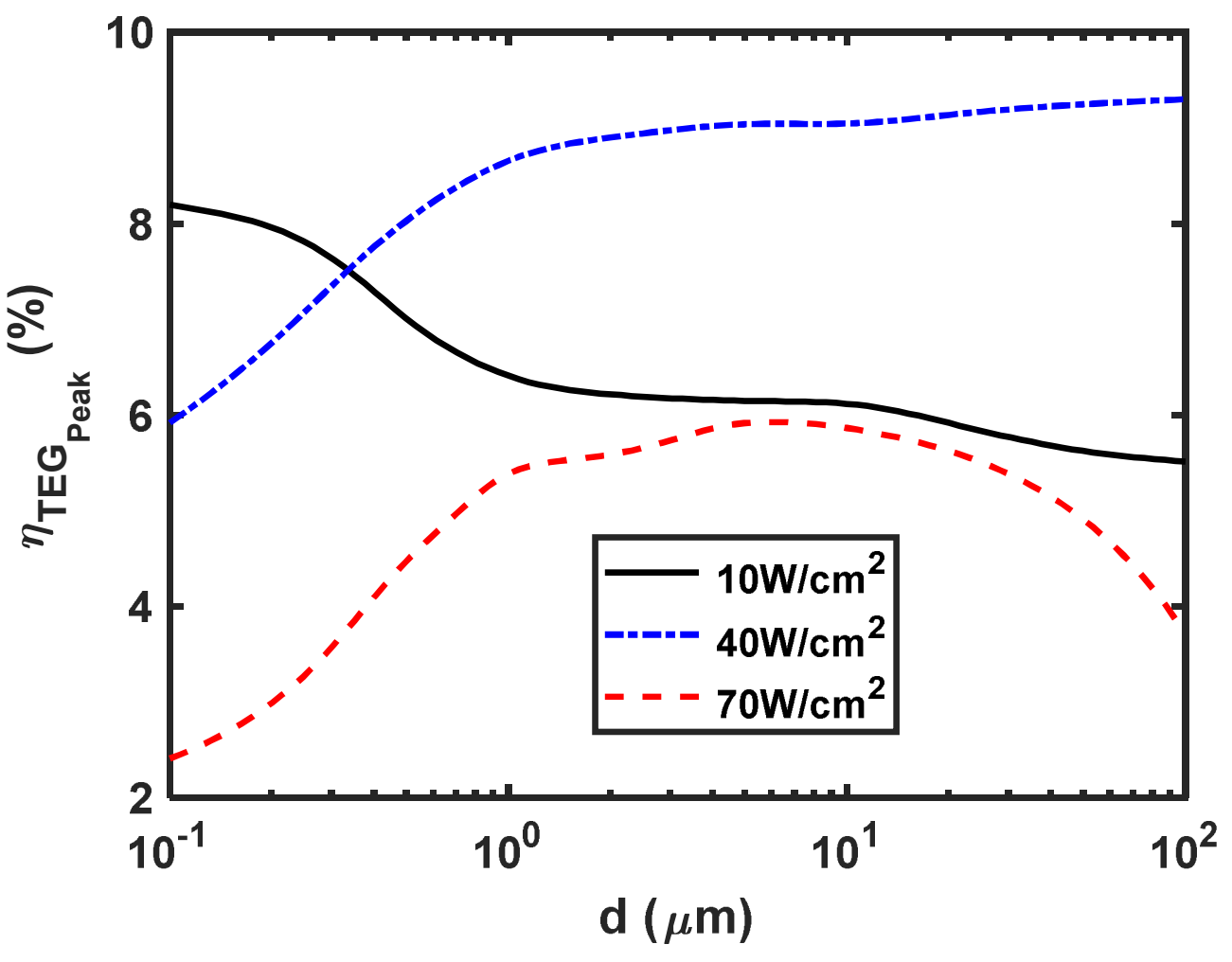

Fig. 8. The efficiency of the TEG bottom stage as a function of the TEC gap width for different concentrated solar irradiances. For each point, the value of $V_{TEC}$ has been chosen to maximize the combined device efficiency.

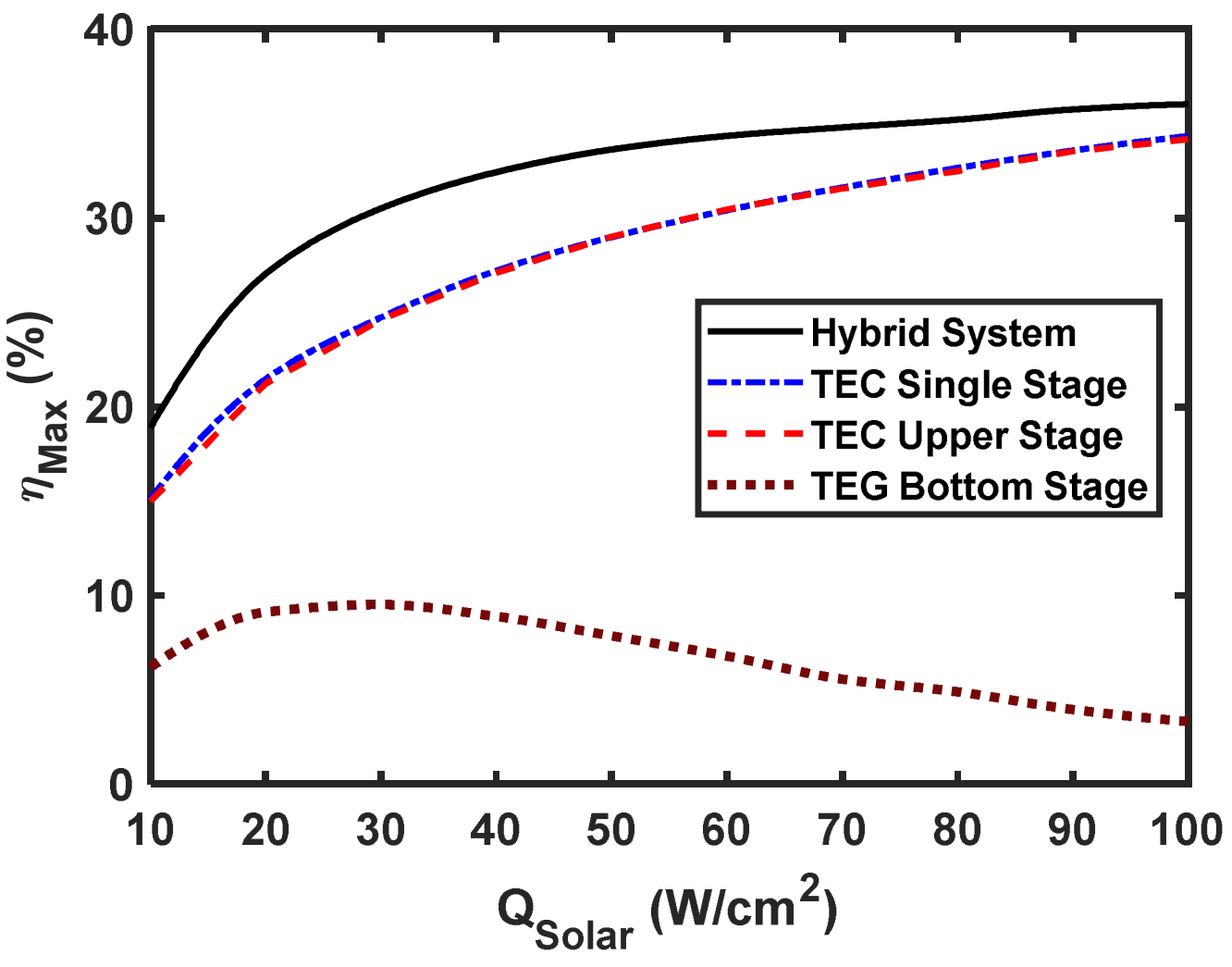

Fig. 9. The maximum efficiency as a function of concentrated solar radiation. The curves shown are for the optimized TEC gap width and output voltage.

The effect of solar radiation intensity on the TEC-TEG hybrid system performance is shown in fig. 9. The efficiency curves for the hybrid system and its different stages are obtained for a TEC voltage and gap width where the combined efficiency of the hybrid system (eq. 5(b)) is maximized. In other words, this efficiency is the maximum point of a three-dimensional graph like fig. 6(c), when plotted for each solar concentration factor. For comparison, the maximum efficiency of a single-stage TEC is also shown in fig. 9. The single-stage TEC operation was simulated by removing the TEG bottom stage and connecting the TEC collector to a heat sink. The maximum efficiency of this stand-alone TEC can then be obtained by optimizing the gap width and operating voltage. One interesting observation from fig. 9 is that the TEG bottom stage has a negligible impact on the performance of the TEC upper stage. In other words, replacing the heat sink with a TEG does not significantly change the maximum efficiency achievable form the TEC for a fixed solar concentration. This is a fortunate outcome as adding a TEG can provide additional electrical power and gain in overall system efficiency without compromising the performance of the TEC. Therefore, despite its low efficiency, a TEG is an excellent candidate for recovering heat that is generated as a by-product of other conversion processes. Such heat would otherwise be released to the environment and wasted. Analysing the efficiency vs intensity trends of the two heat engines, it can be seen that TEC efficiency increases with the solar intensity although at a reduced rate at higher solar concentration. This results from multiple factors including increased external radiation loss, increased back emission from the collector and a more dominant space-charge effect. On the other hand, the TEG efficiency at the maximum operating point of the hybrid system shows a trend with a local maximum as explained before. The values of other important parameters at the maximum efficiency point of the hybrid system are given in table 1. It should be noted that we do not put any particular emphasis on the exact values of these parameters as such values might differ based on the properties of particular materials chosen for the TEC electrodes and TEG thermocouple legs. However, the trend of these parameters with the solar intensity would provide a theoretical guideline for the practical design of such a hybrid system. The corresponding efficiency values are also included in the table to help the readers easily correlate between the efficiency and other parameters.

**Table 1:** Important parameters of the TEC-TEG hybrid system at the maximum efficiency point for different solar intensities. Note that $d_l$ and $d_u$ represent the lower and upper bound of the TEC gap range, respectively, for which the hybrid system efficiency is larger than 90% of its maximum value.

| $Q_{\text{Solar}}$ (Wcm$^{-2}$) | $\eta_{\text{Combined}}^{\text{Max}}$ (%) | $\eta_{\text{TEC}}^{\text{Max}}$ (%) | $\eta_{\text{TEG}}^{\text{Max}}$ (%) | $T_E^{\text{Max}}$ (K) | $T_C^{\text{Max}}$ (K) | $T_G^{\text{Max}}$ (K) | $V_{\text{TEC}}^{\text{Max}}$ (V) | $V_{\text{TEG}}^{\text{Max}}$ (V) | $J_{\text{TEC}}^{\text{Max}}$ (Acm$^{-2}$) | $J_{\text{TEG}}^{\text{Max}}$ (Acm$^{-2}$) | $d_{\text{Max}}$ (µm) | $d_l$ (µm) | $d_u$ (µm) |
|---|---|---|---|---|---|---|---|---|---|---|---|---|---|
| **10** | **18.89** | **14.98** | **6.20** | **1404.7** | **410.4** | **305.8** | **0.75** | **1.67** | **2.04** | **0.230** | **2.15** | **0.87** | **7.05** |
| **20** | **27.04** | **21.23** | **9.12** | **1522.7** | **511.5** | **311.4** | **0.84** | **3.22** | **5.00** | **0.360** | **1.96** | **0.76** | **5.72** |
| **30** | **30.50** | **24.60** | **9.52** | **1642.8** | **588.8** | **316.5** | **0.99** | **4.28** | **7.38** | **0.412** | **1.89** | **0.75** | **5.34** |
| **40** | **32.42** | **27.09** | **8.89** | **1751.3** | **651.4** | **321.5** | **1.14** | **4.98** | **9.36** | **0.427** | **1.89** | **0.73** | **4.98** |
| **50** | **33.62** | **29.01** | **7.86** | **1826.5** | **704.5** | **326.6** | **1.23** | **5.44** | **11.72** | **0.424** | **1.89** | **0.73** | **4.64** |
| **60** | **34.34** | **30.45** | **6.79** | **1908.6** | **748.2** | **331.6** | **1.35** | **5.68** | **13.47** | **0.411** | **1.89** | **0.73** | **4.43** |
| **70** | **34.79** | **31.55** | **5.57** | **1929.7** | **794.0** | **337.7** | **1.35** | **5.81** | **16.34** | **0.389** | **1.87** | **0.73** | **4.27** |
| **80** | **35.20** | **32.48** | **4.90** | **2022.5** | **818.4** | **341.5** | **1.49** | **5.82** | **17.26** | **0.374** | **1.87** | **0.71** | **4.13** |
| **90** | **35.74** | **33.52** | **3.96** | **2039.3** | **854.0** | **347.7** | **1.49** | **5.74** | **20.29** | **0.348** | **1.87** | **0.71** | **3.98** |
| **100** | **36.02** | **34.16** | **3.33** | **2105.3** | **879.5** | **352.3** | **1.58** | **5.62** | **21.62** | **0.327** | **1.87** | **0.71** | **3.84** |

A question might arise as to what the optimal solar concentration for the hybrid generator operation is. To answer this question, let us consider some performance criteria of the hybrid system. These criteria should be based upon what additional gain can be obtained from cascading a TEG stage with the TEC. One such criterion could be the gain in the system efficiency (as a percentage of the TEC single-stage efficiency) resulting from the addition of a TEG bottom stage, which is shown in fig.10 (a) for different solar intensities. It can be seen that efficiency gain reaches a maximum value at a particular solar intensity (26.2% at a solar intensity of 17 $Wcm^{-2}$, which is equivalent to a concentration ratio of the order of 170) and then gradually drops. Another criterion could be the power output from the TEG bottom stage. To discuss this criterion, we consider how the power and efficiency of the TEG are related, which is shown in fig. 10(b) at the maximum efficiency point of the hybrid system, for different solar intensities. Interestingly, the maximum power and efficiency of the TEG stage do not occur at the same point. The TEG power density reaches its maximum (2.34 $Wcm^{-2}$) at a concentration factor of the order of 570, which is significantly higher than that for the maximum efficiency gain. To choose between these criteria, we evaluate the sensitivity of the hybrid system efficiency on the solar intensity, which is shown in fig. 10(c). It can be seen that as we increase the intensity, returns in hybrid system efficiency gradually diminish. Considering the manufacturing cost associated with a large focusing system and the fact that the increase rate in hybrid system efficiency significantly reduces at large concentration, operating the TEG at its maximum power density might not be attractive from the economic perspective.

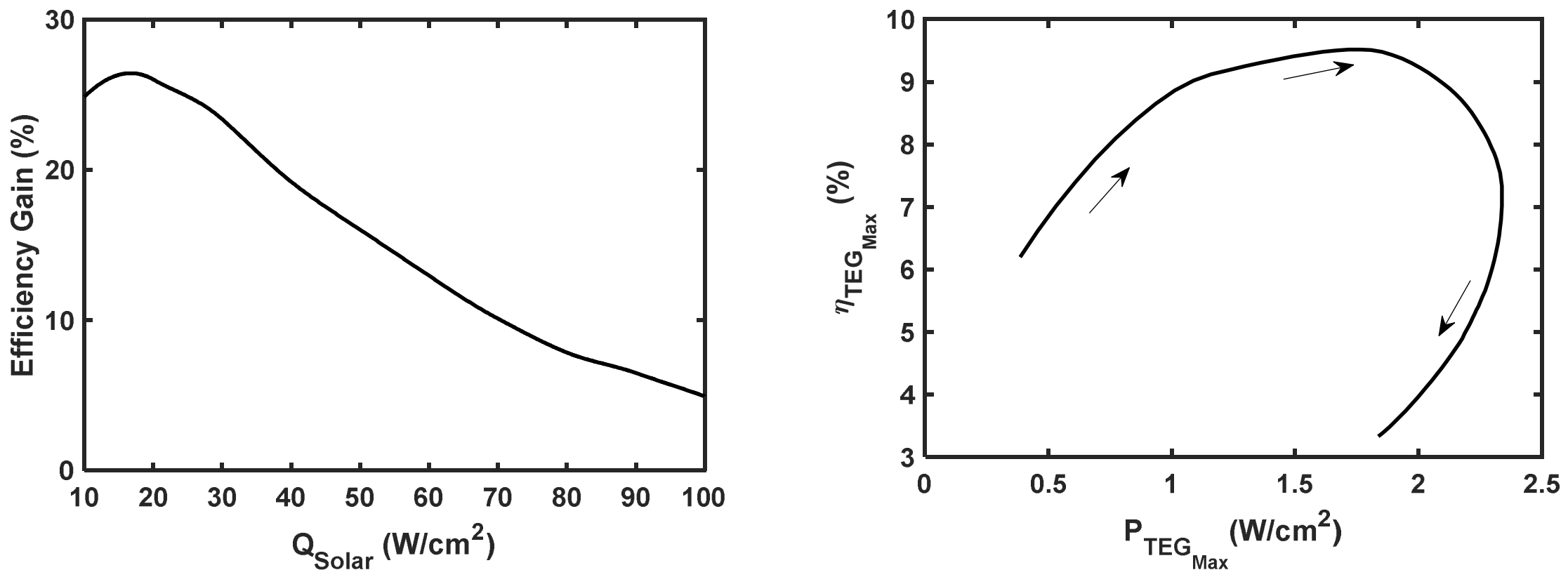

**(a)** **(b)**

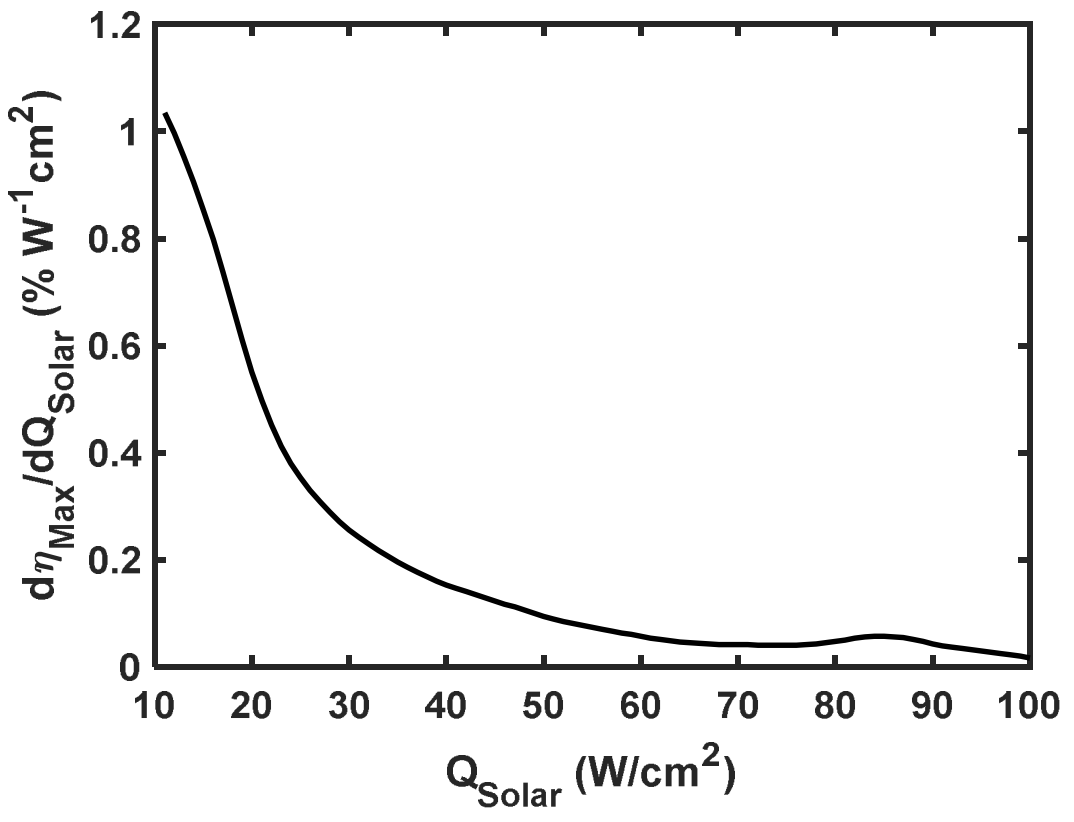

**(c)**

Fig. 10. (a) Gain in maximum efficiency resulting from combined system operation as a function of solar intensity, (b) Efficiency vs power density of the TEG at the maximum efficiency of the hybrid system. Note that the arrows indicate the direction of increasing solar intensity. (c) Efficiency increase rate of the hybrid system as a function of the solar intensity.

## Summary and Conclusion

In summary, this work presented a comprehensive analysis of a TEC-TEG hybrid system. A detailed theoretical model considering the various aspects of the physics of the TEC and TEG heat engines was developed. A self-consistent iterative framework taking into account the energy balance at different electrodes was implemented to explain the operation of the hybrid system under a constant heat flux which, for example, can be due to concentrated solar thermal power. The hybrid system was analysed for wide variations of the TEC output voltage and gap width. The dependence of the TEG performance on the dynamic nature of the waste heat released from the TEC was discussed. The effect of temperature dependent thermoelectric material parameters on TEG operation was illustrated in detail with different examples. The trends of temperatures, power and efficiency with variable TEC gap width and output voltage were also explained. The efficiency of the hybrid system for different concentrated solar intensity and the incremental gain in efficiency from using the TEG bottom stage were evaluated. From these analyses, we conclude the following key findings:

(a) The different energy exchange channels, the temperatures at different parts of the hybrid device, and the currents and output voltages of the TEC and TEG stages all show strong dependences on the TEC interelectrode gap width.

(b) Due to the interplay between the near-field coupling of thermal radiation and space-charge effect, there is an optimal interelectrode gap which results in maximum hybrid system performance. The trend of this optimal gap with input heat flux is relatively flat.

(c) The performance of the TEG stage shows strong temperature dependence. The conversion efficiency and power density of the TEG stage reach their respective maxima at different temperatures, which indicates a trade-off between these two performance metrics of the TEG.

(d) The returns in hybrid system conversion efficiency gradually diminish as the input heat flux density increases. Therefore, in the case of CSP applications, an optimal solar concentration can be found considering the various trade-offs between the hybrid system performance and the costs associated with large concentration factors.

The model thus serves as a powerful tool to understand the operation of a combined TEC-TEG device and design such a system for applications. As future work, this model can be used to study the impacts of different thermionic and thermoelectric materials by carrying out a detailed device performance characterization based on various material parameters. Such a study could help identify optimal materials for TEC-TEG hybrid systems.

**Acknowledgement**

We acknowledge financial support from the Natural Sciences and Engineering Research Council of Canada (RGPIN-2017-04608, RGPAS-2017-507958, SPG-P 478867). This research was conducted, in part, by funding from the Canada First Research Excellence Fund, Quantum Materials and Future Technologies Program. Ehsanur Rahman thanks the Natural Sciences and Engineering Research Council of Canada for a Vanier Canada Graduate Scholarship and the University of British Columbia for an International Doctoral Fellowship and Faculty of Applied Science Graduate Award.

**Nomenclature**

***Alphabets***

| | |
|---|---|
| $A_0$ | Richardson-Dushmann Constant (A $cm^{-2}$ $K^{-2}$) |
| $A$ | Cross-sectional area ($cm^2$) |
| $c$ | Speed of light (m $s^{-1}$) |
| $d$ | Interelectrode distance (μm) |
| $e$ | Electronic charge (coulomb) |
| $\hbar$ | Reduced Planck constant (J s) |
| $I$ | Thermocouple current (A) |
| $J$ | Current density (A $cm^{-2}$) |
| $K$ | Thermal conductance ($WK^{-1}$) |
| $k$ | Thermal conductivity ($Wm^{-1}K^{-1}$) |
| $k_B$ | Boltzmann constant ($m^2$ kg $s^{-2}$ $K^{-1}$) |
| $l$ | Thermocouple length (m) |
| $N$ | Number of thermocouples |
| $P$ | Electrical power density (W $cm^{-2}$) |
| $Q$ | Heat flux density (W $cm^{-2}$) |
| $R$ | Resistance (Ω) |
| $S$ | Coupling coefficients |
| $T$ | Temperature (K) |
| $V$ | Voltage (V) |
| $w$ | Angular frequency (rad $s^{-1}$) |
| $x$ | Position of interelectrode motive (μm) |

***Greek Symbols***

| | |
|---|---|
| $\alpha$ | Seebeck coefficient ($VK^{-1}$) |
| $\beta$ | Parallel wavevector($m^{-1}$) |
| $\varepsilon$ | Effective emissivity |
| $\eta$ | Efficiency (%) |
| $\Theta$ | Mean energy of Plank's oscillator (J) |
| $\lambda_T$ | Wavelength (m) |
| $\mu$ | Thomson coefficient ($VK^{-1}$) |
| $\rho$ | Resistivity ($\Omega$ m) |
| $\sigma$ | Stefan Boltzmann constant (W $cm^{-2}$ $K^{-4}$) |
| $\varphi$ | Work function and interelectrode motive (eV) |

***Subscript***

| | |
|---|---|
| c | Cold side |
| C | Collector/TEG hot side |
| Combined | Value of the hybrid system |
| evan | Evanescent |
| E | Emitter |
| F | Fermi level |
| G | TEG cold side |
| h | Hot side |
| H | TEG hot end |
| *l* | Lower bound at 90% of the hybrid system maximum efficiency |
| L | TEG Cold end; Load |
| Loss | External radiation loss |
| m | Maximum |
| Max | The maximum value of the hybrid system and the corresponding values of the different stages at an optimized TEC voltage and interelectrode distance |
| n | N-type semiconductor |
| p | P-type semiconductor |
| prop | Propagating |
| Peak | Maximum value at an optimized TEC voltage |
| Rad | Interelectrode radiation |
| Solar | Incident solar intensity |
| T | Heat Carried by Thermionically emitted electron |
| TEC | Thermionic converter |
| TEG | Thermoelectric converter |
| u | Upper bound at 90% of the hybrid system maximum efficiency |
| 0 | Ambient |

***Operator***

| | |
|---|---|
| – | Average |